# A nanoCryotron comparator can connect single-flux quantum circuits to conventional electronics

Qing-Yuan Zhao[1], Adam N. McCaughan[1], Andrew E. Dane[1], Karl K. Berggren[1*], and Thomas Ortlepp[2]

[1]Massachusetts Institute of Technology, Department of Electrical Engineering and Computer Science, Cambridge, MA, 02139

[2]CiS Research Institute for Microsensor Systems GmbH, 99099 Erfurt, Germany.

*berggren@mit.edu

**Abstract:**

Integration with conventional electronics offers a straightforward and economical approach to upgrading existing superconducting technologies, such as scaling up superconducting detectors into large arrays and combining single flux quantum (SFQ) digital circuits with semiconductor logic and memories. However, direct output signals from superconducting devices (*e.g.*, Josephson junctions) are usually not compatible with the input requirements of conventional devices (*e.g.*, transistors). Here, we demonstrate the use of a single three-terminal superconducting-nanowire device, called the nanocryotron (nTron), as a digital comparator to combine SFQ circuits with mature semiconductor circuits such as complementary metal oxide semiconductor (CMOS) circuits. Since SFQ circuits can digitize output signals from general superconducting devices and CMOS circuits can interface existing CMOS-compatible electronics, our results demonstrate the feasibility of a general architecture that uses an nTron as an interface to realize a "super-hybrid" system consisting of superconducting detectors, superconducting quantum electronics, CMOS logic and memories, and other conventional electronics.

**Introduction**

Superconducting detectors and electronics are applied in fields where extremely critical specifications are required. For example, superconducting detectors have state-of-the-art sensitivity and are widely used in astronomical observation and quantum information processing[1–3]. In addition, the energy cost and computation speed of superconducting digital circuits are several orders of magnitude better than conventional semiconductor circuits[4]. However, due to the lack of a complete family of electronics, it is difficult and expensive to have an all-superconducting system, such as a large-scale superconducting detector array or a fully-functional microprocessor. On the other hand, semiconductor technology produces an abundance of electronics, such as 100 GHz fast photodiodes and gigabyte-size memories. Moreover, emerging technologies, such as integrated photonics, ferroelectrics, and biological devices are being combined with conventional CMOS technology to achieve hybrid systems of extended functions. Unfortunately, these advanced electronics are rarely compatible with superconducting devices because of the difference in operating temperatures, signal levels, and fabrication processes. As a result, the implementation of superconducting electronics is usually limited to small-scale applications, such as a single-element photon detector.

Future development of superconducting electronics requires a combination of analog and digital circuits to detect, digitize, and post-process data. A straightforward and economical approach is integrating superconducting devices and existing non-superconducting electronics[5–8]. In particular, conventional concerns of temperature compatibility are alleviated by the use of developed cryogenic instruments with higher cooling capability and optimized semiconductor devices whose performances are not degraded at cryogenic temperatures [9,10]. Figure 1 shows the general architecture of such a hybrid system. The output of superconducting electronics, *e.g.*, output pulses from single-photon detectors, is first digitized by a single-flux-quantum (SFQ) circuit, which is a superconducting digital circuit known for its ultralow power dissipation and ultrafast clock frequency [11]. The SFQ chip then communicates with a CMOS circuit, which can not only interface with conventional electronics but can also feedforward results back to the SFQ circuits and the superconducting detectors. This "super-hybrid" system can be implemented for various applications, such as integrating SFQ readouts with superconducting detectors to achieve a large-sized array[12], interfacing SFQ circuits to optical diodes for fast data communication[13], combining SFQ circuits with low-power dissipated cryogenic ferroelectric/CMOS memories[7], and constructing a superconducting quantum/classical computer[14,15].

To close the flow of information loop shown in Fig. 1a, an interface device or circuit between the SFQ circuits and the CMOS circuits is critically required due to the incompatible operation principles between SFQ and CMOS circuits. SFQ circuits use Josephson junctions (JJs), two-terminal devices governed by a gauge-invariant phase, as the basic switching components. The binary logic bits are carried based on whether a single flux quantum ($\Phi_0 = 2\times10^{-15}$ Wb) is trapped in an inductive superconducting loop, or equivalently, whether the loop stores a circulating current. In comparison, CMOS logic circuits are built from transistors, which are three-terminal devices governed by the mobility of carriers. Discrete voltage levels on a capacitor are used to define the binary logic bits. The output pulse generated from switching a JJ has a total energy of $\sim 10^{-19}$ J, which is usually below the sensitivity of a transistor for logic operation[11]. In addition, the duration of the output pulse is only several picoseconds and falls outside of the bandwidth of typical CMOS circuits. Moreover, the output impedance of a switched JJ is only several ohms, which is too low to efficiently drive current into the gate of a transistor with a much higher input impedance (*e.g.*, 50 Ω). Traditional mitigation strategies include using a driving circuit based on modified JJs to magnify the signal amplitude and transform the impedance and speed; examples of such driving stages are those based on superconducting quantum interference device (SQUID) amplifiers[16,17] or latched Josephson junctions (Suzuki stacks[7,18]). Although a CMOS-compatible output can be achieved when multiple JJs are cascaded to increase the signal level and impedance, these driving stages demand a large layout area, have unacceptably high power dissipation, and are intolerant of fabrication variation, thereby limiting their scalability in integration.

Here, we demonstrate a nanoscale driving stage using a single nanocryotron (nTron) for SFQ circuits to access external CMOS electronics. The nTron is a superconducting three-terminal device whose channel superconductivity can be switched by the injection of hot quasiparticles generated at the gate[19]. Unlike JJs, switching out of the superconducting state in a nanowire is followed by the growth of a resistive domain with several kΩ resistance, providing a high output impedance and a higher voltage signal[20,21]. For nTrons made from niobium nitride (NbN) thin film, it takes ~100 ps for the resistive domain to grow from the superconducting state up to several kΩ, providing ~GHz operation speed for driving the load, matching the typical clock frequency of a CMOS digital circuit. In addition to these output operation characteristics, we have demonstrated that the nTron is sensitive to SFQ pulses, making it a promising device for interfacing SFQ circuits with CMOS circuits. Moreover, the active area of the nTron, where the state switches between the superconducting state and the resistive state, is about 0.1 μm$^2$, three orders of magnitude smaller than a Josephson junction in SFQ circuits[22]. Therefore, for future on-chip integration, the nTron-based interface will show incremental scalability to other JJ-based driving circuits.

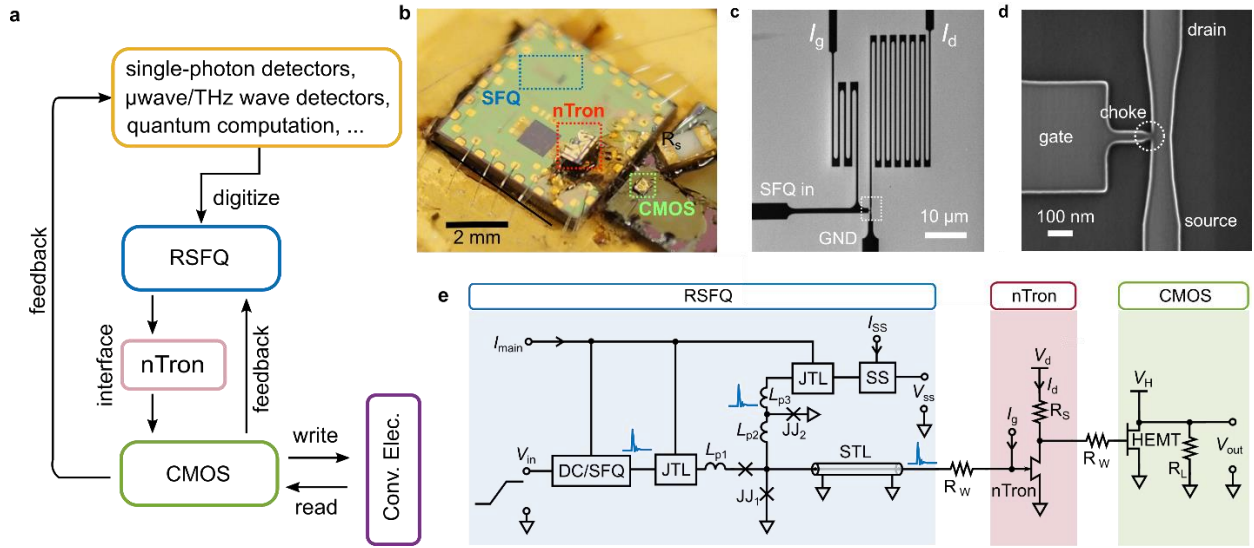

**Figure 1. Connecting superconducting electronics and conventional electronics with nanoCryotron comparators.** (**a**) Blocks show the primary units in a super-hybrid system, where the superconducting electronics (detectors and computation circuits) can access conventional electronics (Conv. Elec.) in a universal and economical architecture. (**b**) Picture of the hybrid circuit used for demonstrating signal flow from an SFQ chip to a CMOS chip through an nTron. (**c**) Scanning electron microscopy (SEM) image of the nTron chip. Configuration of the ports is shown in the figure. The meander wires on the bias ports for the gate and channel are used as choke inductors to supply constant DC currents. **(d)** SEM image of the three terminals around the active area of the nTron, which the dashed box in (c) includes. **(e)** Simplified circuit schematic of the hybrid circuit. The meanings of the abbreviations are: JTL (Josephson transmission line), JJ (Josephson junction), SS (Suzuki stack), STL (superconducting transmission line), $R_w$ (resistance of the bonding wire), and HEMT (high electron mobility transistor).

## Results

In this paper, we demonstrate that an nTron has the sensitivity to read SFQ pulses and that its output is able to drive a commercial transistor. The experiments were conducted on a hybrid system consisting of an SFQ circuit for generating an SFQ pulse, an nTron interface, and a transistor. In this hybrid system, we have demonstrated correct signal transfer of a single flux quantum pulse from an SFQ chip to an off-the-shelf high electron mobility transistor (HEMT, standing in for the front-end of a CMOS circuit), implying the feasibility of a universal approach to integrating superconducting logic with non-superconducting electronics.

### Operation principle of an nTron

An nTron is a three-terminal planar device consisting of a gate nanowire and a channel nanowire with its two ends referred to as drain and source. Its geometry is shown in Fig. 1c and d. The intersection between the gate and the channel is referred as the choke, whose width is designed to be 40 nm. The choke is placed perpendicularly to where the channel is the narrowest and the current density is the highest to maximize the sensitivity of the nTron. Unlike operation of JJs in SFQ circuits, breaking the superconductivity in a nanowire leads to the diffusion of hot quasiparticles and a growth of resistance, which gives classical amplification of the output signal and impedance. As the heat is localized within a sub-100-nm-wide wire, the initial hotspot will expand to a resistive strip of ~ 1μm long, corresponding to a total resistance up to several kilo-ohms in a film with a high sheet resistance (400 Ω/square) in the normal state. When the nTron channel becomes highly resistive, the current that was initially biased through it is expelled out to drive the load, which could be a transistor or a wire capacitor in a memory array.

Recovery of the nTron requires both the local temperature being cooled down to the ambient temperature and the circuit states decaying back to the static levels. Depending on the tradeoff between thermal cooling and current restoration, which is also referred as the electrothermal feedback in superconducting nanowire devices for photon detectors[20,23], the recovery of an nTron can be classified into two different operation modes: the self-reset mode and the latched mode. The current dynamics can be characterized by $\tau_e = L_{ch}/R_L$, where $L_{ch}$ is the inductance of the nTron channel and $R_L$ is the load resistance. For long $\tau_e$, the current through the nanowire recovers slowly, so that the Joule heating from the resistive area is overwhelmed by the cooling to the substrate, enabling the resistive area to switch back to the superconducting state. As a result, after outputting a sharp pulse, the nTron can reset by itself through a process we refer to as "self-reset". The nTron operates in a latched mode for short $\tau_e$, during which time the current rises quickly enough to create an equilibrium between the Joule heating and cooling, supporting a sustained resistive domain[24]. Consequently, once the nTron fires, an active quenching of the bias current is required to reset the device for the next SFQ pulse.

Figure 2a shows the current-voltage (I-V) characteristic of the nTron channel, from which the maximum voltage jump and maximum output resistance in the case of an open load can be derived. When the current increases over the switching current (32.3 µA), the nTron channel switches to a resistive state of 10 kΩ with a voltage jump of 0.15 V. Unlike a JJ whose maximum voltage jump is limited by the energy gap of the superconductor electrodes, the maximum voltage jump in an nTron is governed by the biasing impedance, which is the resistance $R_s$ of the current source in the DC measurement. The hysteresis of the I-V curve shows two current plateaus, corresponding to the self-heating currents for the 200 nm wide channel (9 µA) and the 400 nm wide leads (16 µA). Operation modes can be tuned from the latched mode to the self-reset mode by reducing the output load resistance or increasing the channel inductance. Figure 2b shows the simulation results of nTron output pulses at different $R_L$ with an nTron SPICE model. In the self-reset mode, the output voltage exponentially decays to zero with a time constant of $L_{ch}/R_L$ and the output voltage is proportional to $I_b \times R_L$. Speaking generally, the self-reset mode is promising for detecting SFQ pulses whose arrival time is not synchronized while the latched mode is suitable for driving a highly resistive load. In the following measurements, we operate the nTron in the self-reset mode when driving a 50 Ω load room-temperature amplifier and in the latched mode when driving the HEMT operating at cryogenic temperature.

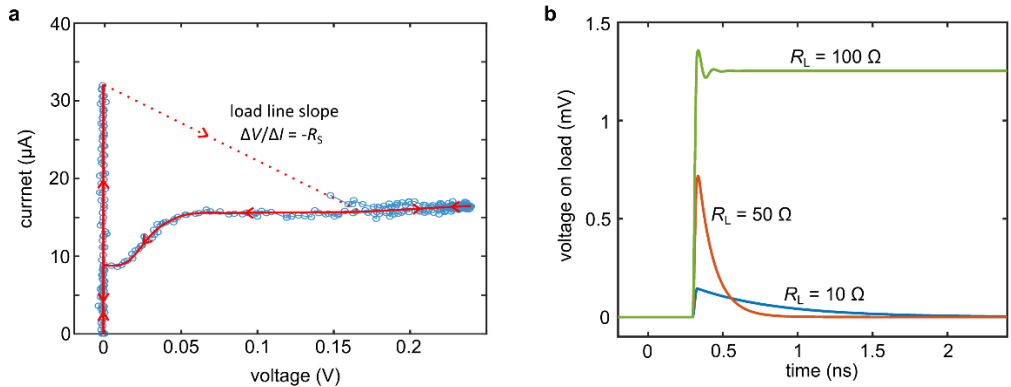

**Figure 2 Switching characteristics of the nTron.** (**a**) DC measurement of the current-voltage relationship of the nTron channel when the gate is open. We applied a sinusoidal voltage $V_d$ with a serial resistor $R_S$ = 10 kΩ to supply current through the nTron channel. The arrows show the hysteresis of the I-V curve. (**b**) SPICE simulation of the output voltage on the load of different resistance $R_L$. The nTron channel is biased at $I_d$ = 20 μA.

**A hybrid SFQ-nTron-CMOS circuit**

Figure 1b shows the hybrid digital system we used for characterizing the nTron interface. It consisted of an SFQ circuit as an SFQ pulse generator, an nTron driving stage for pulse amplification and impedance transformation, and a commercial HEMT used for representing the front-end of a CMOS circuit. External circuits, such as a memory unit for data storage and an optical link for high-speed data transfer, can communicate with this hybrid digital system after the HEMT to complete the hybrid architecture shown in Fig. 1. We connected the individual components with short aluminum bonding wires and immersed them in liquid helium, thus achieving an operation temperature of 4.2 K. A simplified circuit diagram is shown in Fig. 1e.

The SFQ chip was used as an SFQ pulse generator whose circuit details were reported in ref. 25. Its front-end was a DC/SFQ unit, whose purpose was producing a controlled number of SFQ pulses on the leading edge of a square pulse $V_{in}$, which we compared with the nTron output pulses. By adjusting the height and the edge time of the pulse, as shown in Fig.3a and Fig. 4a, the number of SFQ pulses $n$ and interval time between subsequent SFQ pulses was tuned. To export the SFQ pulses off the chip, the generated pulses propagated through a Josephson transmission line (JTL) to switch junction $JJ_1$ ($I_c$ = 225 µA and bias current was 175 µA), whose output pulses then split into two: one triggered a standard Suzuki stack for checking the correct generation of the SFQ pulse sequence and the other output went through a superconducting passive transmission line (PTL) to the nTron gate.

Figures 1c and 1d show the geometry of the nTron, which was designed to have SFQ-level sensitivity. The width of the choke was 40 nm wide and the width of the narrowest part of the channel where the choke intersected was 200 nm. At 4.2 K, the switching current of the channel was $I_{chsw}$ = 32.3 µA and the switching current of the gate was $I_{gsw}$ = 7.4 µA. The choke was connected to the narrowest part of the channel so that the current density was the highest, facilitating the switching of the channel with fewer hot quasiparticles. Taking into account the inductance from the nTron gate to the nTron source (~3 nH), the estimated induced current from an SFQ pulse input to the nTron gate was 0.7 µA. In future integration, by using low inductive connections such as niobium wires and by avoiding bonding wires, the induced current can be increased.

We used an off-the-shelf gallium-arsenide HEMT (Model No. FHX45X, Eudyna) to represent the front-end of a CMOS circuit. The output of the nTron was connected to the gate of the HEMT. We biased the HEMT at $V_{ds\_H}$ = 0.5 V and $I_{ds\_H}$ = 25 mA, where $V_{ds\_H}$ was the voltage between the drain and the source of the transistor and $I_{ds\_H}$ was the current through the drain to the source. At 4.2 K, the transconductance was 0.13 S, corresponding to a voltage gain of 6.5 (16 dB) for a 50 Ω load.

**nTron reading SFQ pulses**

We first demonstrated that an SFQ pulse was able to switch the nTron. In this measurement, we disconnected the HEMT and operated the nTron in the self-reset mode. A room-temperature amplifier (MITEQ-AM-1309, 50 dB gain, 1 kHz ~ 1 GHz) was connected after the nTron to increase the voltage level for observing output pulses with an oscilloscope. The nTron channel was biased constantly at 18 μA. In the hybrid circuit, the resistance and inductance between the SFQ chip and the nTron gate were minimized to maximize the induced current from an SFQ pulse. However, this strong coupling configuration also induced DC current leakage from the SFQ chip to the nTron chip and vice versa, producing difficulties in preserving an isolated biasing current to the nTron gate. As the biasing current of the SFQ chip was much higher than the required bias current to the nTron gate, we found that by slightly adjusting the SFQ bias current $I_{main}$, the leakage current from the SFQ chip was able to efficiently bias the nTron gate for triggering input SFQ pulses. An optimization of the coupling between the SFQ chip and the nTron chip would be required to investigate the dependence of the sensitivity of the nTron on the gate bias; however, this work would be beyond this paper's scope of demonstrating the operation principle of an nTron interface, and would need future on-chip integration to ultimately eliminate parasitic parameters.

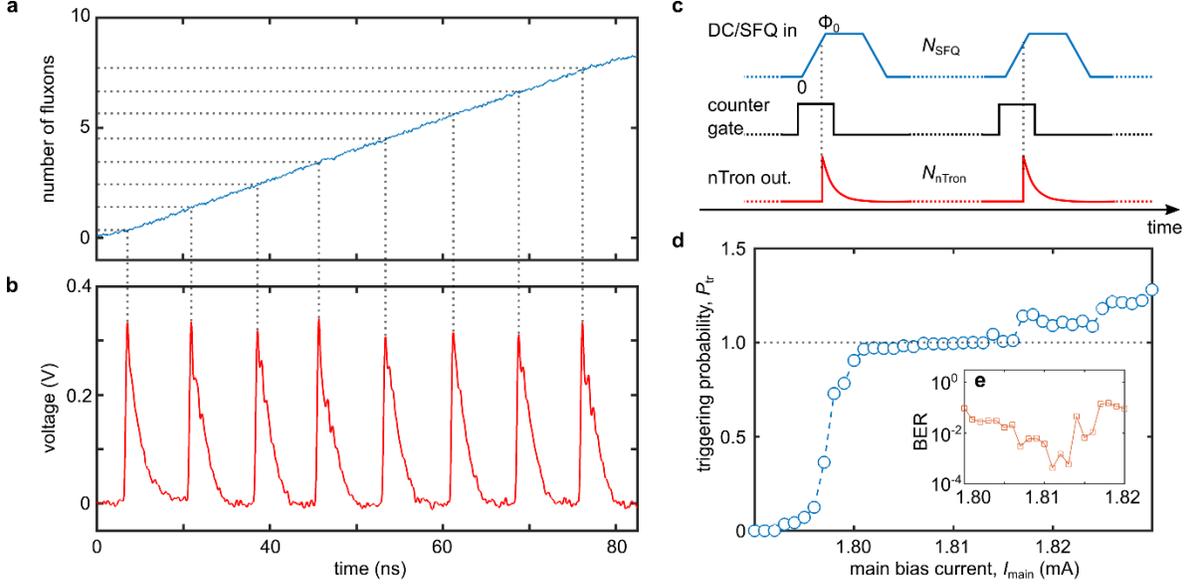

**Figure 3 nTron sensing SFQ pulses.** (**a**) A ramping signal to the DC/SFQ block for generating SFQ pulses. By tuning the amplitude and the rising edge of the signal, multiple SFQ pulses were generated at the traces indicated by the dashed lines. The y-axis was converted to the number of fluxons from $L_p \times I_{in}/\Phi_0$, where $L_p$ = 3.5 pH was the loop inductance in the DC/SFQ circuit and $I_{in}$ was the input current. (**b**) Output pulses from the nTron read from a 50 dB room temperature amplifier. (**c**) Diagram showing the method of measuring the triggering probability $P_{tr}$, defined as the count ratio of the output nTron pulses ($N_{nTron}$) to the input SFQ pulses ($N_{SFQ}$). (**d**) The measured $P_{tr}$ versus the main bias of the SFQ circuits. (**e**) The bit-error-rate (BER) calculated from $|1 - P_{tr}|$.

Figure 3 shows the nTron output triggered by eight SFQ pulses. The load for the nTron was set by the 50 Ω amplifier used for read-out. The amplitude of the eight peaks was 324 mV±7mV. Taking into account the gain of the amplifier, the amplitude of output voltage from the nTron was ~1 mV. Continuously reading out SFQ pulses indicated the SFQ-level sensitivity of the nTron. We defined the triggering probability $P_{tr}$ as the probability that a single SFQ pulse was able to trigger an nTron pulse. To measure $P_{tr}$ we adjusted the height of the DC/SFQ input to have only one SFQ pulse generated at one input pulse and counted the number of input SFQ pulses $N_{SFQ}$ and output nTron pulses $N_{nTron}$, which is shown in Fig. 3c. The triggering probability was then calculated by $P_{tr} = N_{nTron}/N_{SFQ}$.

As previously discussed, by slightly adjusting the main bias current of the SFQ chip $I_{main}$, the DC current through the nTron gate could be tuned, resulting in a dependence of $P_{tr}$ on $I_{main}$. As shown in Fig. 3d, when $I_{main}$ increased from 1.79 mA to 1.81 mA, $P_{tr}$ changed from 0 to 1. As $I_{main}$ kept increasing, one SFQ pulse triggerred the nTron multiple times, resulting in a $P_{tr}$ larger than 1. The bit-error rate (BER) was derived by $E_{tr} = |1-P_{tr}|$; the minimum $E_{tr}$ we measured was $4\times10^{-4}$. Reducing the bit-error rate requires better design of the interface circuit between the SFQ output and the nTron input. However, the experimental results in the hybrid circuit successfully showed that the direct output from an SFQ chip was able to trigger an nTron.

**nTron driving transistors**

Reading nTron pulses with a 50 Ω amplifier demonstrated that the nTron can directly drive 50 Ω loads. To demonstrate the nTron's driving capacity with kΩ loads, we left the gate port of the transistor open and biased the nTron with a voltage source in series with a resistor $R_s$ = 1 kΩ, giving an effective load of $R_s$ for the nTron to drive. As was mentioned in previous sections, for such a highly resistive load, the nTron operated in the latched mode and an AC bias was required to reset the nTron after every SFQ trigger. As shown in Fig. 4, the nTron read an SFQ pulse and subsequently drove the HEMT to output an amplified pulse. The direct output pulse of the HEMT had an amplitude of ~47 mV. Taking the voltage gain of the HEMT (×5), the output voltage from the nTron was derived to be 9.4 mV. The response time, defined as the time constant of the falling edge of the HEMT output shown in Fig. 4d, was 1.5 ns.

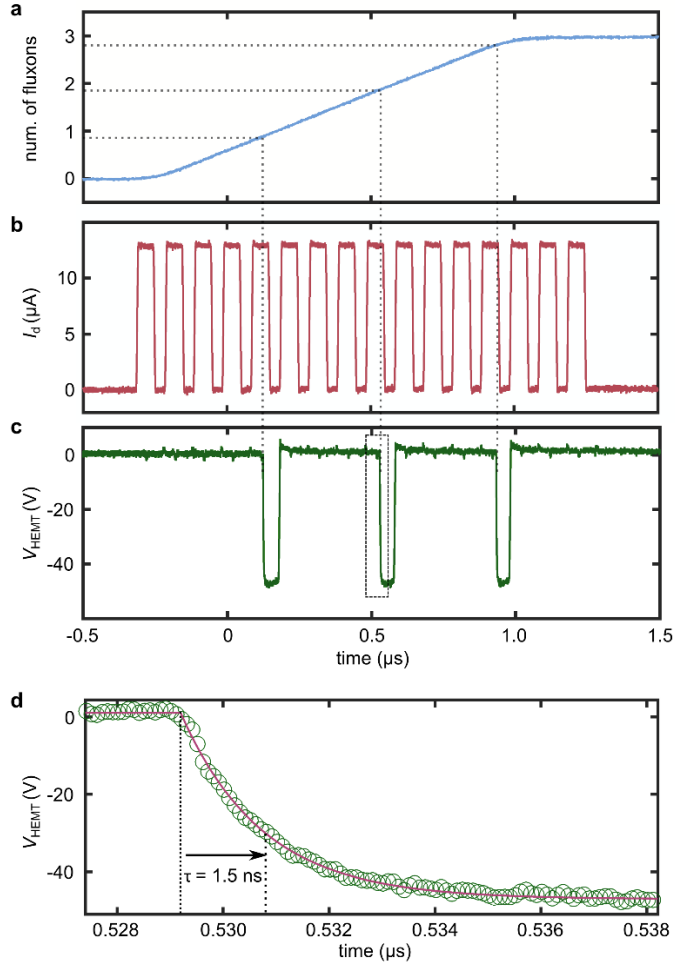

**Figure 4 nTron reading SFQ pulses and driving a transistor.** (**a**) The ramping signal to the DC/SFQ block to trigger SFQ pulse trains. The y-axis is converted to the number of fluxons by the same method described in Fig. 3a. (**b**) AC bias to the nTron channel to operate it in the latched mode. The repetition frequency was 10 MHz. (**c**) Direct output voltage from the HEMT. The falling edges were triggered by the nTron pulses while the rising edges were due to the reset of the AC bias. (**d**) Response time of the transistor measured from the falling edge enclosed in the dashed box shown in (c). We fit the falling edge with an exponential function with a time constant of 1.5 ns, which was used to define the response time of the transistor.

## Discussion

Since short bonding wires were used to connect separate chips, unavoidable parasitic parameters prevented us from fully characterizing the bias margin and grey zone of an nTron comparator. However, we used a hybrid circuit to successfully demonstrate that the nTron has SFQ-level sensitivity and transistor-driving capacity. Under the existing influence from parasitic parameters, the minimum measured BER of switching the nTron was $10^{-4}$, including possible magnetic noise due to an unshielded environment. The high-impedance driving capacity of the nTron was demonstrated by driving a commercial HEMT with an open gate. We have not measured the power dissipation in an nTron, which in principle depends on the geometry of the nanowires and the operation mode. From the SPICE simulation, in the self-reset mode, as the energy initially stored in the inductance of the channel wire is used for creating the normal resistance, the power dissipation for every firing event is proportional to $\frac{1}{2}L_{ch}I_{B\_ch}^2 = 8\times10^{-18}$ J ($L_{ch}$ = 50 nH is the total inductance of the channel wire and $I_{B\_ch}$ = 18 µA is the bias current through the channel). Taking a switching rate of 1 GHz, the power dissipation of a single nTron should be ~8 nW, which is compatible with SFQ circuits. Thus, a number of nTrons could be integrated with RSFQ circuits as interfacing stages on the same chip.

Optimization of nTron comparators for interfacing SFQ circuits with CMOS circuits requires future work in designing a coupling circuit for efficiently transforming SFQ pulses to the nTron gate and an isolated DC bias to the nTron gate. Integration of nTrons with SFQ circuits also requires a compatible fabrication process that can be added to the standard JJ processes. The use of nTron offers a flexible approach to interfacing SFQ circuits with non-superconducting electronics. With its nanoscale geometry, SFQ sensitivity, near GHz switching speed, and kΩ output impedance, the nTron is a promising device for the development of largescale super-"hybrid systems" with unprecedented timing and efficiency.

**Acknowledgements**

The authors thank Emily Toomey for helpful scientific discussion and proof reading of the manuscript. This research was supported by the Office of the Director of National Intelligence (ODNI), Intelligence Advanced Research

**Author Contributions**

K.B., Q.-Y. Z. and T. O. conceived the project. Q.-Y. Z. designed and fabricated the nTrons. T. O. supported the SFQ chip. A. D. supported the superconducting films. Q.-Y. Z., A.M. and T. O. took the measurements. Q.-Y. Z. analyzed the data and wrote the paper with input from K. B. and T. O.. K.B. supervised the project.